# 5G Quality of Service in Bangkok and Metropolitan Areas: Revisiting BTS Skytrain Station Areas

**Therdpong Daengsi[1], Pakkasit Sriamorntrakul[1], Surachai[2] Chatchalermpun, Kritphon Phanrattanachai[3]**

[1]Sustainable Industrial Management Engineering, Faculty of Engineering, Rajamangala University of Technology Phra Nakhon, Thailand
[2]Department of Computer Engineering, Faculty of Engineering, South - East Asia University, Thailand
[3]Department of Computer Engineering, Faculty of Agricultural & Industrial Technology, Phetchabun Rajabhat University, Thailand

| Article Info | ABSTRACT |
|---|---|
|  | This article compares two of the leading mobile network operators in Thailand's telecom market in terms of the service quality of Thailand's 5G networks. The following three factors, download speed, upload speed and latency, which are frequently considered to be indicators of the quality of Internet networks, were examined. The researchers employed the test results to determine an average grade of service that was reached by comparing newly collected data to data that had previously been examined utilizing the same format and application in the middle of May 2021. The typical upload speed dropped from 62.6 Mbps in 2021 to 52.0 Mbps in 2023, while the latency increased from 14.9 to 23.3 milliseconds on average. It was established that the results delivered considerably enhanced quality values despite the fact that the test region in this study only comprised BTS stations. Furthermore, this was the case despite the fact that the test area in this study only encompassed a small percentage of the total population. |

*Corresponding Author:*

Kritphon Phanrattanachai
Department of Computer Engineering, Faculty of Agricultural & Industrial Technology,
Phetchabun Rajabhat University 83 M.11 T. Sadiang A. Muang P. Phetchabun, Thailand 67000
Email: kritphon.ai@pcru.ac.th

## 1. INTRODUCTION

The fifth generation of mobile telecommunications, often known as 5G, is currently the most prevalent type of technology used in the field of telecommunications. It can provide high speed, higher bandwidth, high stability of connections, and extreamly low latency compared to 4G [1]. In present, the deployment of 5G, which has become popular in many countries, including several nations in the ASEAN region as well as Thailand, has taken place. The introduction of this novel technology in Thailand was an important turning point in the country's long and illustrious history of telecommunications. In the first quarter of 2020, one of the leading mobile network operators (MNOs) in the Thai telecommunications industry, which was also the first winner of the frequency spectrum auction, made the official launch of 5G services. This was followed by the launch of 5G services by the second winner of the frequency spectrum auction a few months later [2-3]. Following the conclusion of the auction for the use of the frequency spectrum, the following events took place in the following frequency bands: n3 (1800 MHz), n28 (700 MHz), n41 (2600 MHz), and n258 (26 GHz) [4]. It was anticipated that by the end of this year, Thailand would have 5G coverage for more than 85% of the country's population, while the percentage of 5G devices was expanding to 15% and the number of 5G subscribers would be over 10 million by the end of 2022 [5]. This was all thanks to the advancement of 5G in Thailand.

The rollout of 5G services began in various places throughout the world around the year 2020. This new technology enables compatibility between 5G and older technologies such as LTE and 3G [6][7]. It does so by supporting both the stand-alone (SA) and Non-standalone (NSA) topologies of 5G [8][9]. In general, it is anticipated that 5G will theoretically deliver significant efficiencies in comparison to 4G (see Figure 1 [10]). Since it is capable of supporting a peak data rate of 20 Gbps for the downlink and 10 Gbps for the uplink, respectively, while the goal values for the user experienced data rate are, for example, 100 Mbps and 50 Mbps for the downlink and the uplink, respectively [10], it may support a peak data rate of 20 Gbps for the downlink and 10 Gbps for the uplink. However, in Thailand, there is no official report on Quality of Service (QoS), although QoS is influenced by a number of performance measures [11]. This means that there is no credible institution that can be relied upon to compile this information. The three most common quality of service measurements are the download (DL) speed, the upload (UL) speed, and the latency [12]. They are DL and UL for the users, who are highly familiar with the QoS parameters. They require interaction between the user terminal and the Base Transceiver Station (BTS), which are correspondingly configured for downloading and uploading data [12-13]. Download links are typically designed to function at a quicker speed than upload links [14]. Megabits per second (Mbps) are the units used to measure them. The download and upload rates that are theoretically possible with 5G are 10 Gbps and 1 Gbps, respectively [12][15]. Not only downlink and uplink, but also latency, which is very essential because it is one of the critical variables for the next generation of networks and applications (for example, self-driving cars and telesurgery) [12][16-17]. It is possible for future technologies to lessen its impact, but it will never be eliminated [17-18]. Therefore, this element needs to be kept relatively consistent and below a predetermined limit [12][19]. If it isn't, interactive communication (such as VoIP and online gaming) might not work since the latency is too high. It is generally considered to be beneficial for communications if the latency value is less than 150 milliseconds [12][19]. In the past, the multinational corporation Opensignal would publish its reports once a year about the QoS parameters spanning both DL and UL. However, each time the report has been published, it has been regarded with skepticism due to the fact that its methodology has been called into question, and some of the QoS values have appeared to be lower than the findings that have been tested by end-users in Thailand [3]. As a result, this investigation was carried out in order to determine whether or not the findings from this study linked with the download (DL) and upload (UL) speeds, as well as latency, are compatible with the report that was provided by Opensignal or whether or not they differ.

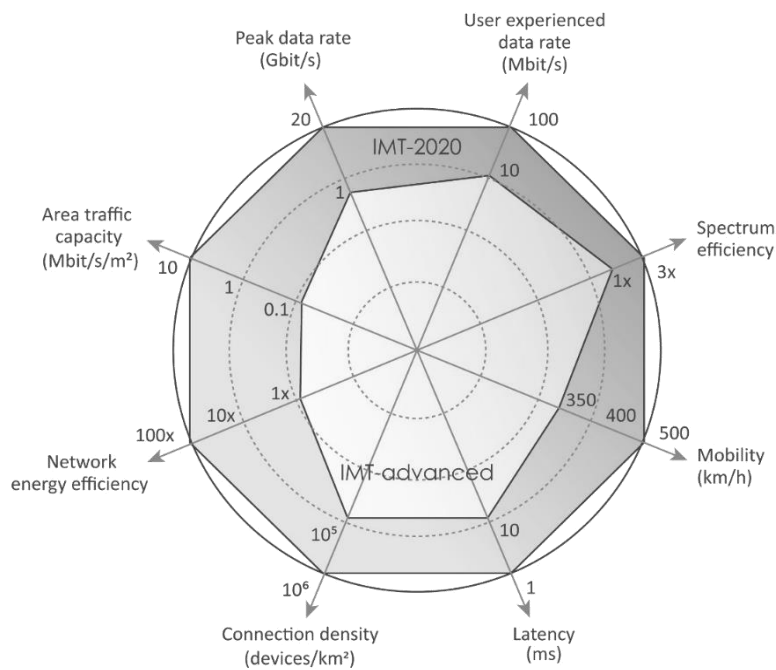

Figure 1. Key capabilities of 5G adopted from [10]

When it comes to the pertinent literature or earlier studies, they can be displayed as shown in Table 1 [3-4] [8] [20-23]. The evaluation of updated 5G performance in Bangkok, Thailand, has not been done in the past, despite the fact that there are a number of tasks linked with 5G performance evaluation. Therefore, for the purpose of evaluating the efficacy of 5G technology in Bangkok using stationary tests [3], the areas surrounding BTS stations were chosen. This is because each station serves between 15,000 and 20,000 passengers on a daily basis. In contrast to [3], only the Opensignal application was used for this study because the purpose of this research was to compare the results to those found in the Opensignal reports [34-35].

The most important contribution that this study made was that it offered the most recent data, which were gathered from actual field tests by utilizing trustworthy methods and technologies. In fact, this study made use of data and methods from [1] and [24]. However, for advancement beyond those earlier works, this study shows the updated results tested in 2023 about 5G performance in the main locations in Bangkok, Thailand, and it has been found that the results are worse than the results tested in 2021. Additionally, this study shows that the results studied are better than the results reported by Opensignal [23].

Table 1. Listerature reviews, adopted from [3]

| Ref. | Quality value of related services | | | | | | Other parameters | Network | | Test mode | | | Tools | Country / Note |
|------|------------|----------------|--------------|---------|--------|------|------------------|----|----|---------------|----------|-----|-------|----------------|
| | Throughput | Download Speed | Upload Speed | Latency | Jitter | Loss | | 4G | 5G | Stay in place | Mobility | N/A | | |
| [3] | - | ✓ | ✓ | ✓ | - | ✓ | - | - | ✓ | ✓ | - | | Speedtest | Thai (11 BTS stations) |
| [4] | - | ✓ | ✓ | - | - | - | - | ✓ | ✓ | - | ✓ | | nPerf, Opensignal, Speedtest | Thailanf (Wat Arun) |
| [8] | - | ✓ | ✓ | ✓ | ✓ | ✓ | MOS | ✓ | - | ✓ | - | | MIQ application with crowdsourcing | Thailand (Nationwide) |
| [20] | ✓ | ✓ | - | ✓ | - | - | - | - | ✓ | ✓ | ✓ | - | Microsoft Azure server | USA (in 3 cities) |
| [25] | - | ✓ | - | ✓ | - | - | Signal Strength, page display Success Ratio and vMOS | ✓ | - | - | - | ✓ | Huawei proprietary tools | Malaysia (rural areas in 3 States) |
| [26] | ✓ | - | - | - | - | - | - | - | ✓ | ✓ | - | | Huawei equipment | Indonesia |
| [21] | - | ✓ | ✓ | - | - | - | - | - | ✓ | - | - | ✓ | speedtest.cn | 105 cities in China |
| [27] | ✓ | - | - | ✓ | - | - | RSSI | - | ✓ | ✓ | - | | Surveillance Task Outline | United Kingdom of Great Britain and Northern Ireland (Backcountry) |
| [28] | - | - | - | ✓ | - | - | RSRP | - | ✓ | - | ✓ | - | Not specified | Republic of Finland |
| [29] | ✓ | - | - | - | - | - | RSRP, SNR | - | ✓ | ✓ | ✓ | - | iPerf | Japanese |
| [30] | ✓ | ✓ | ✓ | - | - | - | - | - | ✓ | ✓ | ✓ | - | Not specified | Japanese |
| [31] | ✓ | ✓ | ✓ | ✓ | - | ✓ | - | ✓ | ✓ | - | ✓ | - | Python scripts and PC on the car | Republic of Finland |
| [32] | ✓ | ✓ | ✓ | ✓ | - | ✓ | - | ✓ | ✓ | - | ✓ | - | Not specified | Jarkatar, Indonesia |
| [33] | - | ✓ | ✓ | ✓ | - | - | - | ✓ | ✓ | - | - | ✓ | Not specified | Jarkatar, Indonesia |
| [34] | - | - | ✓ | ✓ | ✓ | - | SNR, BER | ✓ | - | ✓ | - | - | Mobile Phonr | Melaka, Malaysia |
| [35] | - | - | ✓ | ✓ | ✓ | - | - | ✓ | ✓ | - | - | ✓ | Not specified | Indonesia |
| [36] | - | - | ✓ | - | - | - | - | - | ✓ | - | ✓ | - | Tele-operated driving | Indonesia |

## 2. METHOD

This study, in contrast to many others that came before it, centered its attention on the DL, UL, and latency efficiency of the 5G networks that were provided by two MNOs. The areas surrounding BTS Skytrain station areas were selected for this study in accordance with [3][24]. In this year (2023), the study was condected using the Android smartphone that was a 5G smartphone that had the Mediatek MT6833P Dimensity 810 chipset and the Octa-core (2x2.4 GHz Cortex-A76 and 6x2.0 GHz Cortex-A55) CPU [37]. Dual SIM Dual Standby (DSDS) is supported by it, whereas the Android smartphone that was used in 2021 was the smartphone that had the Kirin 990 5G chipset and the Octa-core (2 x Cortex-A76-based 2.86 GHz + 2 x Cortex-A76-based 2.36 GHz + 4 x Cortex-A55-based 1.95 GHz) CPU [38]. DSDS was supported by it as well. The revisiting study conducted in 2023 involved the use of a smartphone from a different brand and model than the one employed in the previous study. This was necessitated by the unavailability (End of life) of the previous one used in the earlier study at the time of the revisit. Nevertheless, the revisiting study thoughtfully considered and addressed any potential effects arising from differences in chipsets or processors [39]. Both smartphones, of course, came equipped with the Opensignal Speed Test application installed on them [40]. In a manner analogous to [1] and [24], the BTS Skytrain stations were closed down in 2023 due to the fact that those stations are situated on two lines that pass through the most important commercial districts and residential areas in Bangkok and the metropolitan area (see Figure 2 [41]). Daily, there are around 200,000 passengers in total, making up the overall count.

Figure 2. BTS routes in green and light green, adopted from [41]

For the purpose of data collection, stationary tests were carried out at two test points on the platform level and three test points on the concourse level of each BTS Skytrain station (See Figure 2), utilizing 5G unlimited packages from two MNOs that were the first and second winners from the frequency spectrum auction. These tests were carried out within approximately two weeks between March and April 2023, while the older data for comparison were measured by randomly selecting data from the data set that was gathered in May 2021. During this go-around, a total of sixty BTS Skytrain stations were visited again. In addition to that, the findings from the Opensignal reports were incorporated as well [34-35]. Table 2 provides further information regarding the methodology and instruments used. However, the results measured by using the Opensignal application do not show the technology (for example, 4G (LTE), 5G (SA), or 5G (NSA)) while performing the field tests. Because of this, a second application known as the nPerf Speed Test application [42] was utilized in this study as well for checking at each test point before or after conducting each test session using the Opensignal application [40]. It was also utilized in the research conducted in 2021. In the following section, the results of the revisiting, the results of the random pick from 2021, and the findings from Opensignal reports and related results were compared and reported.

Table 2. Measurement Tools

| Item | Detail | |
|---|---|---|
| | 2021 | 2023 |
| Number of stations | 61 | 60 |
| Test date and time | May 10-15, 2021 8:00am-6:00pm | March 20-April 7, 2023 9:00am-1:00pm |
| Number of mobile networks | 2 | |
| Smartphone - 5G Chipset | Kirin 990 | MediaTek Dimensity 810 |
| Test Package | Unlimited from 2 providers | |
| Application used to test | nPerf | |
| Number of test points per station | 4 | 5 |
| Number of data records for this study | 244 | 296 |

Note:   1) There are four stations that have different design compared to other stations, thus the number of test points in the special stations was only four test points.
2) Station N6 was temporary closed, in the revisiting period.

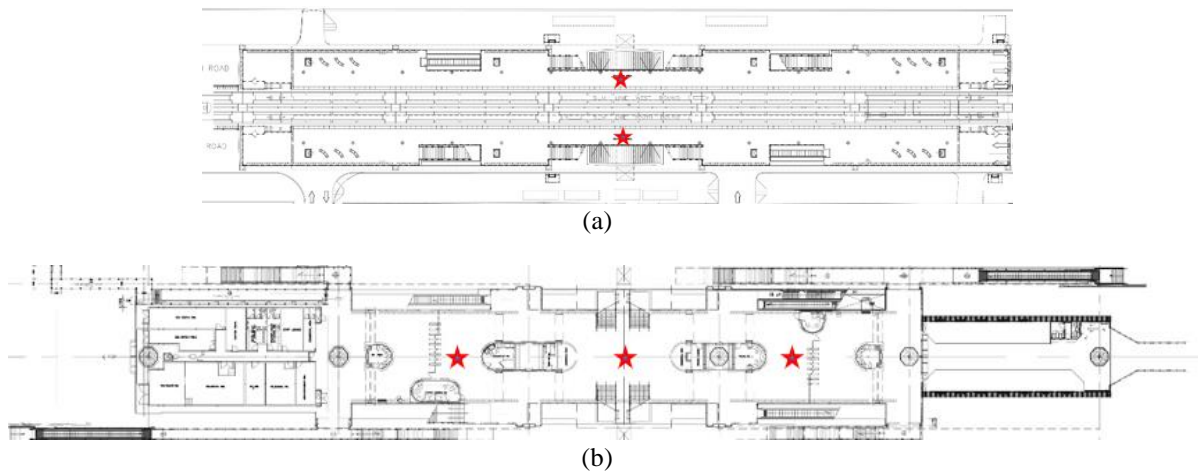

(a)

(b)

Figure 2. Test points at (a) platform level and (b) concourse level

## 3.    RESULTS, ANALYSIS AND DISCUSSION

Following the revisiting and data gathering for the new data in 2023 and the random selection of the data from the old data set collected in 2021 [24]. Those data were processed for getting rid of the outliers, while the results contained within two Opensignal reports presented in 2021 and 2023 respectively were applied [34-35]. The findings obtained from these three distinct sources are then presented and discussed in sections 3.1–3.3.

### 3.1  Download Speed Results

For the DL speeds as shown in Figure 3, it can be described as follows:

-    For overall, the DL speeds provided by MNO1 shows higher performance than the speeds provided by MNO2, both in 2021 and 2023.

-    The average DL speed provided by MNO1 decreased dramatically from almost 310 Mbps in 2021 to 166 Mbps in 2023 approximately.

-    The average DL speed provided by MNO2 increased from almost 85.3 Mbps in 2021 to 144.6 Mbps in 2023. It means that the performance of MNO2 has been improved significantly.

-    The average DL speed from two major MNOs that measured in this study decreased from 196.4 Mbps in 2021 to 140.4 Mbps in 2023.

-    The average DL speed from two major MNOs that obtained from the Opensignal reports [34-35] decreased from 196.4 Mbps in 2021 to 140.4 Mbps in 2023.

-    One can see that the results from this study is consistent with the reports from Opensignal, since the average DL speeds from the studies in 2023 decreased when compared with the average speed measured in 2021.

-    However, for overall the everage DL speeds measured by the team of authors show better results when compared with the resuts inside the Opensignal reports [34-35].

### 3.2  UL Speed Results

For the UL speeds as shown in Figure 4, it can be described as follows:

-    For overall, the UL speeds provided by MNO1 shows better performance than the speeds provided by MNO2, both in 2021 and 2023.

-    The average UL speed provided by MNO1 declined from 83.5 Mbps in 2021 to 71.5 Mbps in 2023 approximately.

-    The average UL speed provided by MNO2 decreased from almost 42 Mbps in 2021 to 33 Mbps approximately in 2023.

-    The average UL speed from two major MNOs that measured in this study decreased from 62.6 Mbps in 2021 to 52 Mbps in 2023.

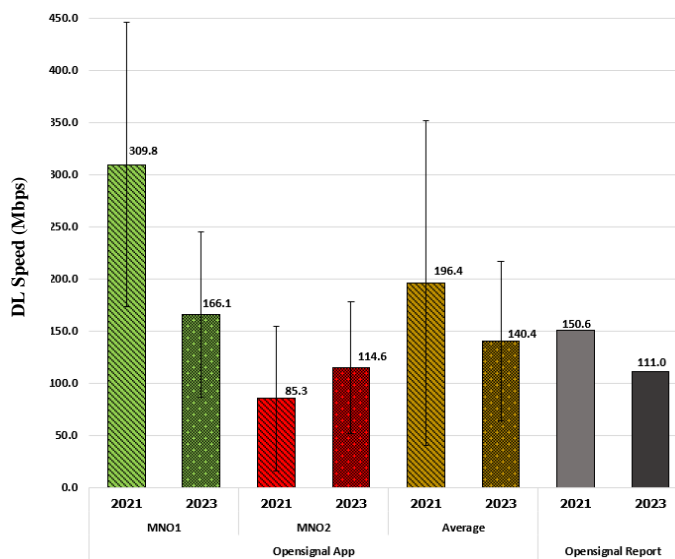

Figure 3. The results of DL speeds

- The average UL speed from two major MNOs that obtained from the Opensignal reports [34-35] slightly declined from 25.0 Mbps in 2021 to 22.2 Mbps in 2023.
- One can see that the results from this study is consistent with the Opensignal reports, since the average UL speeds from the studies in 2023 decreased when compared with the average speed measured in 2021.
- However, for overall the everage UL speeds measured by the team of authors show higher UL speeds compared with the resuts inside the Opensignal reports [34-35].

### 3.3 Latency Results

There is no latency report in the Opensignal reports [34-35], therefore, only the data measured in 2023 and the selected data measured in 2021 were compared in Figure 5, which can be described as follows:

- In 2021 the average latency of 17.3 ms provided by MNO1 shows worse performance than the average latency of 12.5 ms provided by MNO2.
- However, in 2023 the average latency of 22.7 ms provided by MNO1 shows better performance than the average latency of 23.9 ms provided by MNO2.
- For the average latency of 14.9 ms from two MNOs in 2021, it is lower than the average latency of 23.3 ms from MNOs in 2023. It means that the 5G networks measured in 2023 shows worse efficiency than the average value measured in 2021.
- For overall, the trends of latency values for both MNOs are consistent to each other but they are worse when compared to the previous latencies measured in 2021.

### 3.4 Analysis

During the field testing that took place in 2021 and 2023, the nPerf Speed Test program was utilized, and the results relating to the technologies that were served for each test session at each test point were gathered. Table 3 is the appropriate place to present them. From the results in Table 3, it can be described as follows:

- In 2021, it can be implied that the average speeds (DL and UL) provided by MNO1 are higher than MNO2 because the percent of MNO1 (69.3%) can provide higher number of 5G channels (5G NSA) when compared to the percent of MNO2 (61.9%).

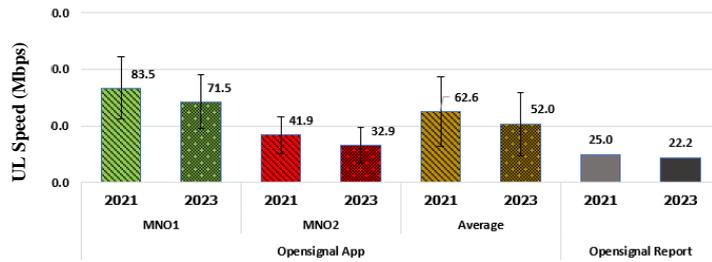

Figure 4. The results of UL speeds

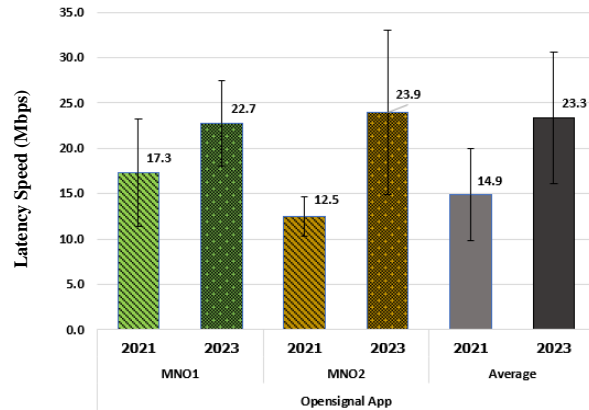

Figure 5. The results of latency values

Table 3. Comparison of the results associated with technology used in each test session from nPerf

| Year | Operator | Service System | | | Estimatation of 5G Coverage Areas |
|------|----------|----------------|---|---|----------------------------------|
| | | 5G (NSA) | 4G (LTE) | Total | |
| 2021 | MNO1 | 169 | 75 | 244 | 169/244*100% = 69.3% |
| | MNO2 | 151 | 93 | 244 | 151/244*100% = 61.9% |
| 2023 | MNO1 | 275 | 21 | 296 | 275/296*100% = 92.9% |
| | MNO2 | 271 | 25 | 296 | 271/296*100% = 91.6% |

- In 2023, it is consistent the percents in 2021, MNO1 can provide more percent of 5G channels (92.9%) than MNO2 (91.6%). However, it is questionable that the percents of 5G NSA in 2023 are higher than the percents in 2021 but the average data rates measured in 2023 are lower than the data rates measured in 2021.

### 3.5 Analysis

From the results, the additional analysis was performed using t-tests, following [4]. The results from each MNO measured in 2021 and 2023 were compared using the six hypotheses below and only the new data between two MNOs measured in 2023 were also compared using additional hypotheses. Whereas comparison between the old data from two MNOs were ignored, since one can see that for overall about data rates in 2021, MNO1 obviously shows better performance than MNO2. All hypotheses are presented as follows:

H1: the average 5G DL speed provided by MNO1 measured in 2023 and the DL speed measured in 2021 are the same or not.

H2: the average 5G DL speed provided by MNO2 measured in 2023 and the DL speed measured in 2021 are the same or not.

H3: the average 5G UL speed provided by MNO1 measured in 2023 and the UL speed measured in 2021 are the same or not.

H4: the average 5G UL speed provided by MNO2 measured in 2023 and the UL speed measured in 2021 are the same or not.

H5: the average 5G latency provided by MNO1 measured in 2023 and the latency measured in 2021 are the same or not.

H6: the average 5G latency provided by MNO2 measured in 2023 and the latency measured in 2021 are the same or not.

H7: the average 5G DL speed provided by MNO1 and the DL speed provided by MNO2 measured in 2023 are the same or not.

H8: the average 5G UL speed provided by MNO1 and the UL speed provided by MNO2 measured in 2023 are the same or not.

H9: the average 5G latency provided by MNO1 and the latency provided by MNO2 measured in 2023 are the same or not.

As shown in Table 3, one can see that all hypotheses show significant differences. Therefore, the analysed results can be used to confirm that the results shown in Figure 3-5 are reliable.

### 3.5 Discussion

After conducting the study and obtaining the results, there are several issues that can be discussed as follows:

- Considering Figure 3-5, the DL and UL speeds measured in 2023 are lower than the DL and UL speeds measured in 2021, while the latencies assessed in 2023 are higher than the measured latencies in 2021 [3][24]. That means recent 5G networks provide lower performance than they did in 2021.

- The causes of lower speeds may be the higher number of 5G subscribers and 5G user equipment (UEs), while the licenses to allow users to access the full capacity of 5G might be limited. Furthermore, business and/or marketing reasons may be a few of the causes of reducing 5G performance since MNOs have to invest in and deploy new 5G equipment and systems to replace the old equipment and systems in order to support 5G users or subscribers.

- Overall, the average 5G performance assessed by the auditors is higher than the performance shown in the Opensignal report.

- It has been found that the results of 5G speeds from this study are consistent with the Opensignal reports. The overall DL and UL speeds are trending to decrease significantly. However, it is inconsistent with DL speeds provided by MNO2, since its 5G DL performance has improved significantly.

Table 4. The Analyzed results from hypothesis tests

| Hypothesis | p-values | Meaning |
|---|---|---|
| H1 | <0.001 | The average 5G DL speed from MNO1 measured in 2023 is worse than the DL speed measured in 2021 significantly. |
| H2 | <0.001 | The average 5G DL speed from MNO2 measured in 2023 is worse than the DL speed measured in 2021 significantly. |
| H3 | <0.001 | The average 5G UL speed from MNO1 measured in 2023 is worse than the UL speed measured in 2021 significantly. |
| H4 | <0.001 | The average 5G UL speed from MNO2 measured in 2023 is worse than the UL speed measured in 2021 significantly. |
| H5 | <0.001 | The average 5G latency from MNO1 measured in 2023 is worse than the latency measured in 2021 significantly. |
| H6 | <0.001 | The average 5G latency from MNO2 measured in 2023 is worse than the latency measured in 2021 significantly. |
| H7 | <0.001 | The average 5G DL speed from MNO1 measured in 2023 is better than the DL speed from MNO2 significantly. |
| H8 | <0.001 | The average 5G UL speed from MNO1 measured in 2023 is better than the UL speed from MNO2 significantly. |
| H9 | 0.044 | The average 5G latency speed from MNO1 measured in 2023 is worse than the latency speed from MNO2 significantly. |

Remark: it is significant if p-value < 0.05 for 95% confidence interval

- As shown in Table 3, although the coverage areas of 5G networks in 2023 in Bangkok are higher than in 2021, they cannot guarantee that they can provide better 5G performance.
- This study covers the BTS Skytrain station areas in Bangkok and metropolitan areas only for the period of time the studies were conducted; therefore, the results from this study are not representative of the 5G performance of the whole of Bangkok or Thailand.
- The smartphones used for the field tests in 2021 and the one used in 2023 are not the same brand and model because, because of a limited budget, the same brand and model as in 2021 were not available in that period, while the smartphones used in 2023 were personal phones that were available in that period of time. The power of chipsets may be different [39] and this issue might be investigated in depth in future work.
- In this revisiting of BTS Skytrain stations, the number of test points increased from one point to two points at the platform level of all stations, except four stations that have different floor plans. The additional test points may impact 5G performance since the platform lavel is approximately 12 meters higher than the ground. Therefore, this issue should be investigated in depth in the future.
- Previously, there were three major MNOs, but two of them have merged together, and only one smartphone was used in the field tests because of budget limitations. Thus, only two major MNOs were evaluated in this study.
- The method using stationary mode for 5G performance evaluation in this study might be applied to other routes of railway systems in Bangkok and the metropolitan area. Furthermore, it can be applied in other countries to evaluate and/or verify whether 5G performances in their countries are consistent with the Opensignal reports or other reports or not.
- This study was performed using stationary tests only; mobility tests should be an option for the next study in the future. While other QoS parameters, such as loss and jitter, have not been considered yet. Therefore, these methods and parameters should be considered for future works.
- This study was mainly based on the Opensignal application, while one interesting feature of the nPerf Speed Test application was also applied. In the future, other applications (e.g., the Speedtest application by Ookla) should be options.

## 4. CONCLUSION

After this study, which looked at three QoS parameters—downlink (DL) speed, uplink (UL) speed, and latency—it was found that the performance of 5G in the field is very different from how it works in theory. According to this study, there will be a significant decline in 5G performance from major MNOs between 2021 and 2023. A larger number of 5G subscribers and 5G UEs, as well as business and/or marketing considerations, are a few explanations for the degraded performance of 5G.

When compared to the 2021 evaluation, the DL speed in 2023 was just 140.4 Mbps, down from 196.4 Mbps. In 2023, the UL speed was measured at 52.0 Mbps, down from 62.6 Mbps in 2021. However, by 2023, the delay had risen to 23.3 ms, up from 14.9 ms in 2021. The field test results match those from Opensignal. Although Opensignal's published values for DL and UL are lower than those found in this investigation, the aggregate results can be used to confirm that the speeds shown here are substantially faster. However, future research should think about other network metrics (e.g., jitter and loss), apps (e.g., Speedtest), and methodologies (e.g., mobility tests).

## ACKNOWLEDGEMENTS

Thanks to Rajamangala University of Technology Phra Nakhon, Phetchabun Rajabhat University and South - East Asia University for supporting this study.

## BIOGRAPHIES OF AUTHORS

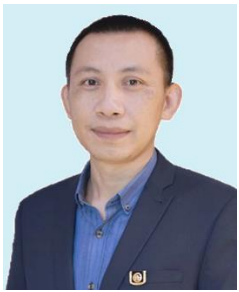

**Therdpong Daengsi**    is an Assistant Professor in the Faculty of Engineering, RMUTP. He received B.Eng. in Electrical Engineering from KMUTNB in 1997. He received a Mini-MBA Certificate in Business Management and M.Sc. in Information and Communication Technology from Assumption University in 2006 and 2008 respectively. Finally, he received Ph.D. in Information Technology from KMUTNB in 2012. He also obtained certificates including Avaya Certified Expert – IP Telephony and ISO27001. With 19 years of experience in the telecom business sector, he also worked as an independent academic for a short period before being a full-time lecturer at present. His research interests include VoIP, QoS/QoE, mobile networks, multimedia communication, telecommunications, cybersecurity, data science, and AI. He can be contacted at email: therdpong.d@rmutp.ac.th.

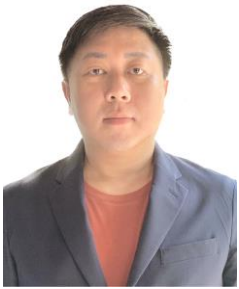

**Pakkasit Sriamorntrakul**    is now a master student in the Faculty of Engineering, Rajamangala University of Technology Phra Nakhon (RMUTP). He received B.Eng. in Computer Engineering from Mahidol University in 2005. He obtained the Avaya Certified Expert Certificate and was the Avaya Certified Support Specialist in IP Telephony. He also held other certificates, including Cisco Certified Network Professional, Microsoft Certified Systems Administrator, and VMware Certified Professional 5. He had 18 years of experience in system, network, and telecom businesses. His research interests include high-performance computer systems and networks, VoIP quality measurement, security, mobile network, AI, and IoT. He can be contacted at email: pakkasit-s@rmutp.ac.th.

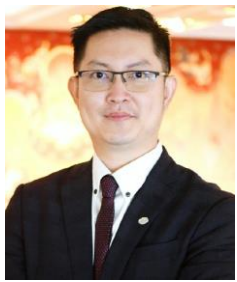

**Surachai Chatchalermpun**    is a special lecturer of Computer Engineering, Faculty of Engineering, South - East Asia University. He recieved first class honor of B.S. degree in computer engineering, King Mongkut's University of Technology Thonburi (KMUTT), Bangkok, Thailand in 2008. He is an expert for Cybersecurity & Data Privacy and Risk Management, he is now a CSPO (Country Cyber Security & Privacy Officer), Huawei Technologies (Thailand) Co., Ltd. He was a CISO at Krungthai Bank, the largest state-own enterprise in Thailand, and he was a regional head of IT security at Maybank in Asia-Pacific. Also, he held many international certificates such as EU-GDPR-CEPAS, CISSP, CDPSE, FIP, CIPP/E, CIPM, CISA, CISM, CSSLP, SSCP, CEH, ECSA, ISO27001 and MIT and Harvard executive certificates. His research interests include cyber security, 5G cyber attack, phishing attack, secure software development life cycle, cloud security, data privacy and data protection. He can be contacted at email: surachai.won@gmail.com

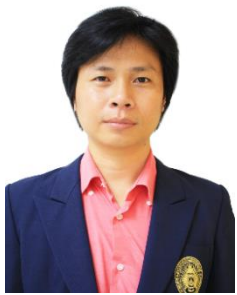

**Kritphon Phanrattanachai**     Kritphon Phanrattanachai is an Assistant Professor in Faculty of Agricultural & Industrial Technology, Phetchabun Rajabhat University (PCRU), Thailand. He received the BSc. degree in electrical industrial from Phetchabun Rajabhat University, Thailand, in 2002. He received the MSc degree in electrical technology from KMUTNB in 2009. and Ph.D. in Tech.Ed. from KMUTNB in November 2019. Also, he is now an assistant president of PCRU. His research interests include circuit synthesis, simulation of linear and non-linear circuits and systems, IoT, QoS/QoE, mobile networks, telecommunications. He can be contacted at email: kritphon.ai@pcru.ac.th.